\documentclass[aps,prd,twocolumn,superscriptaddress,amsmath,showpacs]{revtex4}
\usepackage{graphicx}
\usepackage{dcolumn}
\usepackage{bm}
\usepackage{natbib}
\usepackage{color}

\newcommand{\be}{\begin{equation}}
\newcommand{\ee}{\end{equation}}
\newcommand{\bea}{\begin{eqnarray}}
\newcommand{\eea}{\end{eqnarray}}

\def\neff{N_{\rm eff}}

\newcommand{\Mpc}{{\rm ~Mpc}}

\urldef\mgcamb\url{http://www.sfu.ca/~aha25/MGCAMB.html}

\begin{document}
\mbox{}\\ \hfill CERN-PH-TH/2014-071, LAPTH-027/14
\title{Probing nuclear rates with Planck and BICEP2}

\author{Eleonora Di Valentino}
\affiliation{Physics Department and INFN, Universit\`a di Roma ``La Sapienza'', Ple Aldo Moro 2, 00185, Rome, Italy}

\author{Carlo Gustavino}
\affiliation{INFN, Universit\`a di Roma ``La Sapienza'', Ple Aldo Moro 2, 00185, Rome, Italy}

\author{Julien Lesgourgues}
\affiliation{Institut de Th\'eorie des Ph\'enom\`enes Physiques, EPFL, CH-1015, Lausanne, Switzerland, and\\
CERN, Theory Division, CH-1211 Geneva 23, Switzerland, and\\
LAPTh (CNRS - Universit\'e de Savoie), BP 110, F-74941 Annecy-le-Vieux Cedex, France
}

\author{Gianpiero Mangano}
\affiliation{INFN, Sezione di Napoli, Complesso Univ. Monte S. Angelo, Via Cintia, I-80126 Napoli, Italy}

\author{Alessandro Melchiorri}
\affiliation{Physics Department and INFN, Universit\`a di Roma ``La Sapienza'', Ple Aldo Moro 2, 00185, Rome, Italy}

\author{Gennaro Miele}
\affiliation{Dipartimento di Fisica, Universit\'a di Napoli ``Federico II'' and \\ INFN, Sezione di Napoli, Complesso Univ. Monte S. Angelo, Via Cintia, I-80126 Napoli, Italy}

\author{Ofelia Pisanti}
\affiliation{Dipartimento di Fisica, Universit\'a di Napoli ``Federico II'' and \\ INFN, Sezione di Napoli, Complesso Univ. Monte S. Angelo, Via Cintia, I-80126 Napoli, Italy}

\begin{abstract}
Big Bang Nucleosynthesis (BBN) relates key cosmological parameters to the primordial abundance of light elements.  In this paper, we point out that the recent observations of Cosmic Microwave Background anisotropies by the Planck satellite and by the BICEP2 experiment constrain these parameters with such a high level of accuracy that the primordial deuterium abundance can be inferred with remarkable precision. For a given cosmological model, one can obtain independent information on nuclear processes in the energy range relevant for BBN, which determine the eventual $^2$H/H yield.
In particular, assuming the standard cosmological model, we show that a combined analysis of Planck data and of recent deuterium abundance measurements in metal-poor damped Lyman-alpha systems provides
independent  information on the cross section of the radiative capture reaction $d(p,\gamma)^3$He converting deuterium into helium. Interestingly, the result is higher than the values suggested by a fit of present  experimental data in the BBN energy range ($10 - 300$ keV), whereas it is in better agreement with {\it ab initio} theoretical calculations, based on models for the nuclear electromagnetic current derived from realistic interactions. Due to the correlation between the rate of the above nuclear process and the effective number of neutrinos $\neff$, the same analysis points out a $\neff>3$ as well.
We show how this observation changes when assuming a non-minimal cosmological scenario. We conclude that further data on the $d(p,\gamma)^3$He cross section in the few hundred keV range, that can be collected by experiments like LUNA, may either confirm the low value of this rate, or rather give some hint in favour of next-to-minimal cosmological scenarios.
\end{abstract}


\maketitle

\section{Introduction} \label {sec:intro}

Big Bang Nucleosynthesis (BBN, see e.g. \cite{Iocco1} for a recent overview) offers one of the most powerful methods
to test the validity of the cosmological model around the MeV energy scale.
Two key cosmological parameters enter BBN computations, the energy density in
baryons, $\Omega_b h^2$, and the effective neutrino number, $\neff$, defined such that
the energy density of relativistic particles at BBN is given by

\begin{equation}
\rho_\mathrm{rel}=  \rho_\gamma \left(1+ \frac{7}{8}\, \left( \frac{4}{11} \right)^{4/3} \neff \right) \ , 
\label{eq:neffdef} 
\end{equation} 

\noindent where $\rho_\gamma$ is the Cosmic Microwave Background (CMB) photon energy density, given today by
$\rho_{\gamma,0}\approx 4.8 \times 10^{-34}$ g\,cm$^{-3}$.

Recent measurements of CMB anisotropies obtained by the
Planck satellite are in very good agreement with the
theoretical predictions of the minimal $\Lambda$CDM cosmological model. They 
significantly reduce the uncertainty on the parameters of this model, and provide strong bounds on its possible extensions \cite{PlanckXVI}.
Assuming a given cosmological scenario and standard BBN dynamics,
it is now possible to infer indirectly from Planck data
the abundance of primordial nuclides with exquisite precision.
For example, assuming $\Lambda$CDM, the Planck constraint on the baryon density, $\Omega_b h^2=0.02207\pm0.00027$, can be translated into a prediction for the primordial deuterium fraction using the public BBN code \texttt{PArthENoPE} \cite{parthenope}\footnote{In this paper we use a version of \texttt{PArthENoPE} where the $d(p,\gamma)^3$He reaction rate is updated to the best fit experimental determination (see section \ref{sec:nuc}). The deuterium fraction given by the public version of \texttt{PArthENoPE} is slightly different, but the change in the central value is at the level of 4 per mille, only.}
\be
^2\mbox{H}/\mbox{H}=(2.65\pm 0.07) \cdot 10^{-5} ~~(68\%~{\rm C.L.}) \ ,
\label{deutBBN}
\ee
This constraint is competitive with the most recent and precise direct observations.
Recently, the authors of Ref.~\cite{Cooke:2013cba} (see also \cite{Pettini-Cooke}) presented a new analysis of  all known deuterium absorption-line systems, including some new data from very metal-poor Lyman-alpha systems at redshift $z=$ 3.06726 (visible in the spectrum of the quasar QSO SDSS J1358+6522) and at redshift
$z= 3.04984$ (seen in QSO SDSS J1419+0829). Their result
\be 
^2{\mbox H}/\mbox{ H}=(2.53\pm0.04)\cdot 10^{-5} ~~(68\%~{\rm C.L.})\ ,
\label{pettini}
\ee 
is smaller than the (indirect, model-dependent) cosmological determination from CMB data, but with a comparable uncertainty.

These two deuterium abundance determinations, while broadly consistent, are off by about  two standard deviations. This small tension might well be the result of small experimental systematics, either in Planck or in astrophysical deuterium measurements. However, the point of this paper is to underline that current BBN calculations could also be plagued by systematics in the experimental determination of nuclear rates. As explained in the following, the main uncertainty for standard BBN calculations of $^2$H comes from the rate of the radiative capture reaction $d(p,\gamma)^3$He. A recent review of the experimental status for this process can be found in \cite{Adelberger:2010qa}. The low energy limit of its cross section $\sigma(E)$ (or equivalently, of the corresponding astrophysical factor $S(E)$ \footnote{ We recall that the energy-dependent cross section $\sigma(E)$ is related to the energy-dependent astrophysical factor $S(E)$ through $ \sigma(E)= S(E) e^{-2 \pi \eta}/E$, where $\eta$ is the Sommerfeld factor.}) is well-known thanks to the results of the underground experiment LUNA \cite{casella}. However, during BBN, the relevant energy range in the center of mass is rather around $E \simeq 30 - 300$ keV. For such energies, the uncertainty on the cross section is at the level of 6-10\% when fitting $S(E)$ with a polynomial expression. This translates into a theoretical error on the primordial $^2$H/H ratio of the order of 2\% (for a fixed value of the baryon density and $\neff$), comparable to the experimental error in the above cosmological determination (\ref{deutBBN}) or astrophysical determination (\ref{pettini}).

Recently, a reliable {\it ab initio} nuclear theory calculation of this cross section has been performed in \cite{Viviani:1999us,Marcucci:2004sq,Marcucci:2005zc}. The uncertainty on this prediction can be conservatively estimated to be also of the order of 7\% \cite{Nollett:2011aa}. However, the theoretical result is systematically larger than the best-fit value derived from the experimental data in the BBN energy range. By plugging the theoretical estimate of the cross section in a BBN code one finds that more deuterium is destroyed for the same value of the cosmological baryon density, and thus the predicted primordial $^2$H abundance results to be smaller \cite{Nollett:2011aa}. Interestingly, this could be a way to reconcile the slightly different values of $^2$H/H measured in astrophysical data and predicted by Planck. Indeed, the result quoted in eq. (\ref{deutBBN}) using the public BBN code \texttt{PArthENoPE} \cite{parthenope} relies on a value of the cross section $d(p,\gamma)^3$He inferred from nuclear experimental data (the default value for the $d(p,\gamma)^3$He rate used in the code was calculated in \cite{Serpico:2004gx}, and agrees at the 1.4\% level with the best-fit result of \cite{Adelberger:2010qa}). 

Further data on this crucial cross section in the relevant energy range might be expected from experiments such as LUNA. While waiting for such measurements one can find out to which extent the deuterium measurement of \cite{Cooke:2013cba} can be made even more compatible with Planck predictions when the rate of the reaction $d(p,\gamma)^3$He is treated as a free input parameter. We will address this issue assuming different cosmological models: the minimal $\Lambda$CDM model, $\Lambda$CDM plus extra radiation, a non spatially-flat universe, etc. This simple exercise points out  that, remarkably, present CMB data are powerful enough to provide information on nuclear rates. Moreover, we will see that our results give independent support to the theoretical calculation of \cite{Marcucci:2005zc}. Of course, this close interplay between astrophysical observations and nuclear physics is not new. It is worth while recalling the role that the solar neutrino problem played in the quest for a more accurate solar model, and the impact of this question on experimental efforts for measuring specific nuclear cross sections. 

The paper is organized as follows. In the next section, we discuss in more details the nuclear rates which are most relevant for the determination of the primordial deuterium abundance  and its theoretical error. We introduce a simplified way to parameterize the level of uncertainty still affecting the $d(p,\gamma)^3$He reaction rate, found to be sufficient for our analysis. 
In Section III, we describe our method for fitting cosmological and astrophysical data. We present our results in Section IV, and discuss their implications in Section V.

\section{The primordial Deuterium as function of cosmological parameters and nuclear rates}\label{sec:nuc}

As well known, the theoretical value of the primordial $^2$H/H abundance  is a rapidly decreasing function of the baryon density parameter $\Omega_b h^2$. If we consider a slightly more general cosmological model with extra radiation,  it grows as $\neff$ increases. Finally, this value depends on the cross section of a few leading nuclear processes, responsible for the initial deuterium production and its subsequent processing into $A=3$ nuclei. More precisely, the calculation depends on the thermal rate of such processes, obtained by convolving their energy-dependent cross section $\sigma(E)$  with the thermal energy distribution of incoming nuclei during BBN. The four leading reactions are listed in Table \ref{tablerates}. Note that the uncertainties reported in the Table, like all other results quoted in this paper, unless otherwise stated, are calculated with a version of \texttt{PArthENoPE} where the $d(p,\gamma)^3$He reaction rate is updated to the best fit determination of \cite{Adelberger:2010qa}.

In the past, BBN calculations were based on the experimental determination of the cross section of nuclear processes, measured in laboratory experiments. The situation has changed recently, since detailed theoretical calculations are now available, at least for some reaction. For example, this is the case for the cross section of the neutron-proton fusion reaction $p(n,\gamma)^2\mbox{H}$, for which a very accurate result could be derived using pion-less effective field theory, with a theoretical error below the percent level \cite{Rupak,Chen} (see e.g. \cite{Serpico:2004gx} for further details). Using \texttt{PArthENoPE}, one can propagate this error to the primordial deuterium abundance. The resulting uncertainty is very small, $\sigma_{^2{\mathrm H}/{\mathrm H}}= 0.002 \cdot 10^{-5}$, i.e. of the order of 0.1\% (for $\Omega_b h^2$ fixed at the Planck best-fit value).

The cross sections of  $d-d$ fusion reactions, $d(d,n)^3\mbox{He}$ and $d(d,p)^3\mbox{H}$, are still determined using experimental data. They have been measured in the 100 keV range with a 1-2\% uncertainty \cite{Leonard}. This leads to a propagated uncertainty on the deuterium primordial abundance at most of the order of 1\%, see Table \ref{tablerates}. 

The main source of uncertainty is presently due to the radiative capture process $d(p,\gamma)^3 \mbox{He}$ converting deuterium into helium. The present experimental status for the corresponding astrophysical factor $S(E)$ (where $E$ is the center of mass energy) is reviewed in \cite{Adelberger:2010qa}. As we already mentioned, when fitting a polynomial expression for $S(E)$ to the raw data, now dominated by the LUNA results \cite{casella}, one finds that the uncertainty at 68\% C.L. grows from 6\% in the low energy limit to 19\% around 1~MeV. In the energy range relevant for BBN, the uncertainty is in the range 6-10\%, which gives an error on the primordial deuterium abundance of order $\sigma_{^2{\mathrm H}/{\mathrm H}}= 0.062 \cdot 10^{-5}$, as reported in Table \ref{tablerates}. This uncertainty is comparable to the experimental error estimated by \cite{Cooke:2013cba}, and dominates the error budget. In addition, the best fit value of $S(E)$ inferred from the data in the range $30$ keV$\leq E \leq 300$ keV is lower than the theoretical result of  \cite{Viviani:1999us,Marcucci:2005zc} by about 1$\sigma$. This difference may have an impact on the concordance of Planck results for the baryon density with the deuterium abundance measured by \cite{Cooke:2013cba}. 
\begin{table}
\begin{tabular}{|c|c|c|}
\hline
Reaction & Rate Symbol &  $\sigma_{^2{\mathrm H}/{\mathrm H}} \cdot 10^{5}$ \\ \hline\hline
$p(n,\gamma)^2\mbox{H}$& $R_1$ &  $\pm 0.002$ \\ \hline
$d(p,\gamma)^3 \mbox{He}$& $R_2$ &  $\pm 0.062$\\  \hline
$d(d,n)^3\mbox{He}$ & $R_3$ &  $\pm 0.020$ \\ \hline
$d(d,p)^3\mbox{H}$ & $R_4$ &  $\pm 0.013$\\ \hline
\end{tabular}
\caption{List of the leading reactions and corresponding rate symbols controlling the deuterium abundance after BBN. The last column shows the error on the ratio $^2$H/H coming from experimental (or theoretical) uncertainties in the cross section of each reaction, for a fixed baryon density $\Omega_b h^2=0.02207$.}
\vspace{0.5cm}
\label{tablerates}
\end{table}

Using \texttt{PArthENoPE}  with the best fit experimental cross section for the $d(p,\gamma)^3 \mbox{He}$ reaction, one can check that the best fit value of the astrophysical determination of the deuterium abundance, $^2\mbox{H}/\mbox{H}=2.53  \cdot 10^{-5}$ \cite{Cooke:2013cba}, corresponds to $\Omega_b h^2 = 0.02269$. However, in the case of the minimal cosmological model (i.e. the spatially flat $\Lambda$CDM model, with no extra relativistic species and $\neff=3.046$ \cite{Mangano}), we have seen that Planck data yield  $\Omega_b h^2=0.02207\pm0.00027$~(68\% C.L.). Hence there is a moderate 2$\sigma$ tension, which could be relaxed either by assuming a more complicated cosmological model compatible with higher values of the baryon density, or by adopting the theoretical value of the $d(p,\gamma)^3 \mbox{He}$ cross section \cite{Marcucci:2005zc}. In the latter case, if we stick to the $\Lambda$CDM model, the same range for the baryon density leads to 
\be
^2\mbox{H}/\mbox{H}=(2.58\pm 0.07) \cdot 10^{-5} \  ,
\label{deutBBN2}
\ee
in nice agreement with the astrophysical determination at the 1$\sigma$ level.  In other words, increasing the $d(p,\gamma)^3 \mbox{He}$ thermal rate has the same effect of increasing the cosmological baryon fraction. 

\begin{figure}
\includegraphics[width=.45\textwidth]{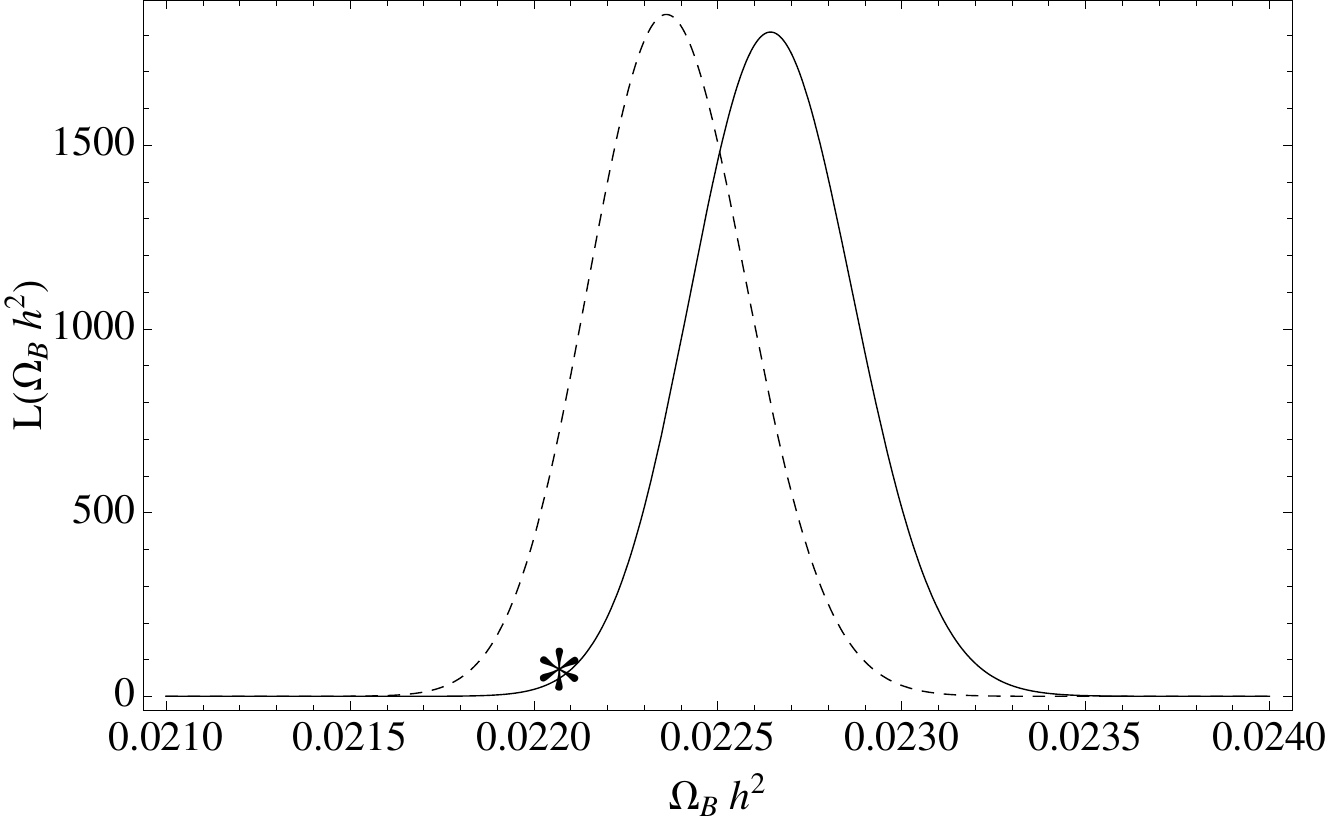}
\caption{\label{fig1} The likelihood L$(\Omega_b h^2)$, assuming the astrophysical determination of the primordial deuterium abundance $^2$H/H by Cooke et al. \cite{Cooke:2013cba}, adopting either the experimental best fit $R_2^{ex}(T)$ (solid) or {\it ab initio} calculation $R_2^{th}(T)$,  (dashed) ~\cite{Marcucci:2005zc}. The star shows the Planck best fit value of $\Omega_b h^2$ in the minimal $\Lambda$CDM model.}
\end{figure}

This is illustrated in Fig.~\ref{fig1} where the likelihood function L($\Omega_b h^2,R_2$)
\bea 
&& \mbox{L}(\Omega_b h^2,R_2) =  \nonumber \\ && \exp \left(- \frac{(^2\mbox{H}/\mbox{H}_{th}(\Omega_b h^2,R_2) -\mbox{$^2$H}/\mbox{H}_{ex})^2}{\sigma^2_{ex} } \right) \ ,  \label{likely0}
\eea
is plotted versus baryon density in two different scenarios. Indices $th$ and $ex$ refer to the theoretical value of $^2$H/H and to the experimental result of \cite{Cooke:2013cba}, respectively. The solid line corresponds to $R_2^{ex}(T)$ obtained by using the best fit of experimental values for the $d(p,\gamma)^3 \mbox{He}$ cross section, while the dashed line relies on the theoretical prediction of the same cross section~\cite{Marcucci:2005zc}, whose corresponding rate is denoted by $R_2^{th}(T)$. The latter brings the agreement with the Planck $\Lambda$CDM value of $\Omega_b h^2$ from the 2$\sigma$ to the 1$\sigma$ level. 
Note that, in calculating those likelihoods, we only included the experimental error on astrophysical measurements of the deuterium fraction,$\sigma_{ex}= 0.05$. Indeed, our purpose is to show what the baryon probablility could like after a future measurement campaign of the $d(p,\gamma)^3 \mbox{He}$ astrophysical factor, assuming a small uncertainty and two different central values for this measurement.
If the theoretical calculation of ~\cite{Marcucci:2005zc} was experimentally confirmed, the likelihood profile would shift  to the dashed curve.

In the next section, we will generalize this study to non-minimal cosmological scenarios. The aim is to see whether, by combining CMB and BBN data, we can grasp some robust information on the value of the thermal rate $R_2$ {\it preferred} by cosmology. To this end, it is enough to parametrize the generic $R_2(T) $ in terms of an overall rescaling factor $A_2$, namely $R_2(T) = A_2 \, R_2^{ex}(T)$,  and use it  in \texttt{PArthENoPE}. This approximation may sound too simplistic, but one can easily check that the ratio $R_2^{th}(T)/R_2^{ex}(T)$ is almost independent of temperature in the region relevant for BBN. For example using a constant rescaling factor $A_2=1.055$ one can mimic $R_2^{th}(T)$ with quite a good precision, and this conclusion holds for any value of $\Omega_b h^2$ in the range from 0.021 to 0.024, with at most a 0.2\% difference in the predicted deuterium abundance.  Hence, the use of a constant rescaling factor $A_2$ is reliable enough for our purpose, and offers the advantage of limiting the number of extra free parameters to one.

Assuming this ansatz, we introduce the baryon likelihood function, L($\Omega_b h^2,A_2$), through
\bea 
&& \mbox{L}(\Omega_b h^2,A_2) =  \nonumber \\ && \exp \left(- \frac{(^2\mbox{H}/\mbox{ H}_{th}(\Omega_b h^2,A_2) -\mbox{$^2$H}/\mbox{H}_{ex})^2}{\sigma^2_{ex} +\sigma^2_{th}
} \right) \ ,  \label{likely}
\eea
where the theoretical value  is a function of the baryon density and the $d(p,\gamma)^3 \mbox{He}$ thermal rate rescaling factor $A_2$, and again we use the experimental value and its squared uncertainty, see Eq. (\ref{pettini}). Finally, $\sigma^2_{th}$ is the squared propagated error on deuterium yield due to the present experimental uncertainty on $R_2$.

\section{Data analysis method}\label{sec:data}

Our main dataset consists in the Planck public data release of March 2013 \cite{PlanckXV}, based on Planck temperature completed by WMAP9 polarization at low $\ell$. We also consider the recent B modes
polarization data (5 bins) from the BICEP2 experiment \cite{bicep2}.  
We combine these two CMB datasets (referred as Planck+WP and Planck+WP+BICEP2 respectively) with the deuterium abundance likelihood function L($\Omega_b h^2,A_2$) (referred as BBN).

Occasionally, we will also include the direct measurement of the Hubble constant by \cite{hst} (referred as HST),  and information on Baryon Acoustic Oscillations by SDSS-DR7 at redshift $z = 0.35$ \cite{sdss-dr7}, by SDSS-DR9 at $z = 0.57$ \cite{sdss-dr9}, and by WiggleZ at $z = 0.44$, $0.60$, $0.73$ \cite{wiggle-z} (referred alltogether as BAO).

For the data analysis method, we will use indifferently the publicly available Monte Carlo Markov Chain packages {\sc CosmoMC} \cite{Lewis:2002ah} (\url{http://cosmologist.info/cosmomc/}) and {\sc Monte Python} \cite{Audren:2012wb} (\url{http://montepyhton.net}), which rely on the Metropolis-Hastings algorithm for exploring the parameter space, and on a convergence diagnostic based on the Gelman and Rubin statistics. We use the latest version of the two codes (April 201a), which include the support for the Planck Likelihood Code v1.0 (see \url{http://www.sciops.esa.int/wikiSI/planckpla/}) and implement an efficient sampling of the parameter space using a fast/slow parameter decorrelation \cite{Lewis:2013hha}.
We checked that the results from the two codes were identical. To evaluate the deuterium abundance produced during the Big Bang Nucleosynthesis, we use the \texttt{PArthENoPE} code, minimally modified in order to account for the global rescaling factor $A_2$.

We will first consider the Planck+WP dataset assuming the minimal $\Lambda$CDM model with six free parameters: the density of baryons and cold dark matter $\Omega_{ b}h^2$ and $\Omega_{ c}h^2$, the ratio $\theta$ of the sound horizon to the angular diameter distance at decoupling, the optical depth to reionization $\tau$, the amplitude $A_S$ of the primordial scalar fluctuation spectrum at $k=0.05\Mpc^{-1}$, and the spectral index $n_S$ of this spectrum. We extend this list of free parameters to include the rescaling factor $A_2$, affecting only the determination of the primordial deuterium abundance. For this model, we consider purely adiabatic initial conditions, we impose spatial flatness, we fix the effective number of neutrinos to its standard value $N_{\rm eff}=3.046$~\cite{Mangano}, and we consider the sum of neutrino masses to be $0.06$eV as in the \cite{PlanckXVI}.

Subsequently, we will study several extensions of the minimal $\Lambda$CDM model, with extra free parameters: the neutrino effective number $N_{\rm eff}$, the spatial curvature of the universe parametrised by $\Omega_k=1-\Omega_c-\Omega_b-\Omega_{\Lambda}$, and the amplitude of the lensing power spectrum $A_{\rm L}$~\cite{Calabrese}.

Finally, we consider a $\Lambda$CDM+r framework where we allow the possibility for a gravitational
wave background with tensor to scalar amplitude ratio $r$. In this case we include the BICEP2
dataset, assuming the B mode signal claimed by this experiment to be the genuine signature of
primordial inflationary tensor modes.
Since the amplitude of tensor modes measured by BICEP2 is in tension with the upper limit 
on $r$ coming from the Planck experiment, we also consider two further extensions that could in principle solve the tension: an extra number of relativistic particles parametrized by $N_{\rm eff}$ 
(see e.g. \cite{giusarma14}) and a running of the spectral index $dn_S/dlnk$ \cite{bicep2}.

\section{Results}\label{sec:results}

In Table \ref{table0}, we report our results for the parameters of the minimal $\Lambda$CDM model (plus the nuclear rate parameter $A_2$ and the derived cosmological parameter $H_0$), using the data combinations Planck+WP+BBN and PLANCK+WP+BBN+BAO. 

\begin{table*}
\begin{center}
\begin{tabular}{|c|c|c|}
\hline\hline
Parameter & Planck+WP & Planck+WP \\
 & +BBN & +BBN+BAO \\
\hline
$\Omega_bh^2$ &$0.02202\pm0.00028$ &$0.02209\pm0.00025$ \\
$\Omega_{\rm c}h^2$ &$0.1200\pm0.0026$ &$0.1188\pm0.0017$ \\
$\theta$ &$1.04129\pm0.00063$ &$1.04144\pm0.00058$ \\
$\tau$ &$0.089\pm0.013$ &$0.091\pm0.013$ \\
$n_s$ &$0.9599\pm0.0073$ &$0.9625\pm0.0058$  \\
$\log[10^{10} A_s]$ &$3.089\pm0.025$ &$3.089\pm0.025$ \\
$H_0 [\mathrm{km}/\mathrm{s}/\mathrm{Mpc}]$& $67.2\pm1.2$ &$67.74\pm0.78$\\
$A_2$ &$1.155\pm0.082$ &$1.138\pm0.076$ \\
\hline\hline
 \end{tabular}
 \caption{Constraints on cosmological parameters (at the $68 \%$ confidence level) in the case of the minimal $\Lambda$CDM model.}
 \label{table0}
 \end{center}
 \end{table*}

\begin{figure}
\includegraphics[width=.35\textwidth]{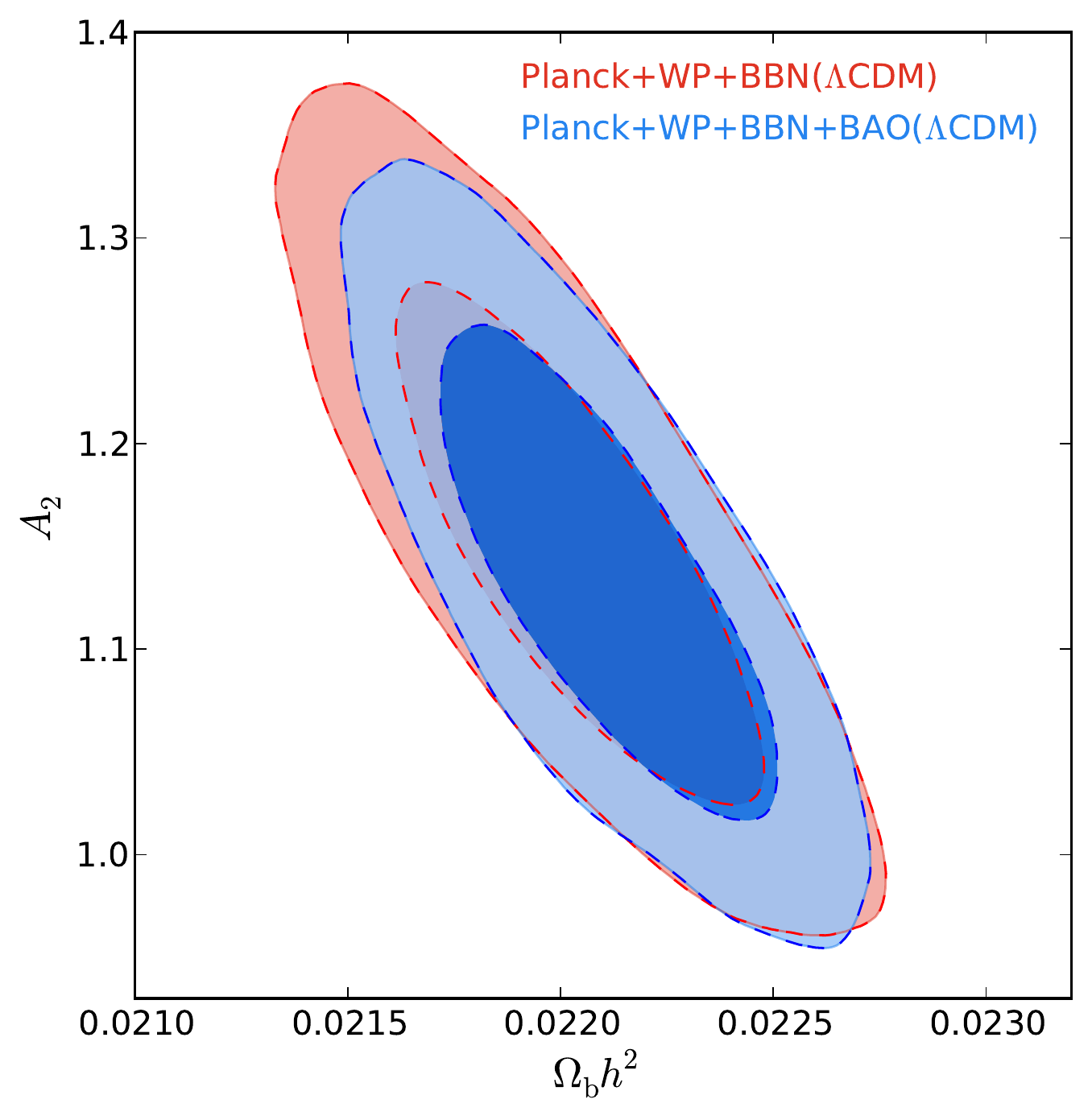}
\includegraphics[width=.35\textwidth]{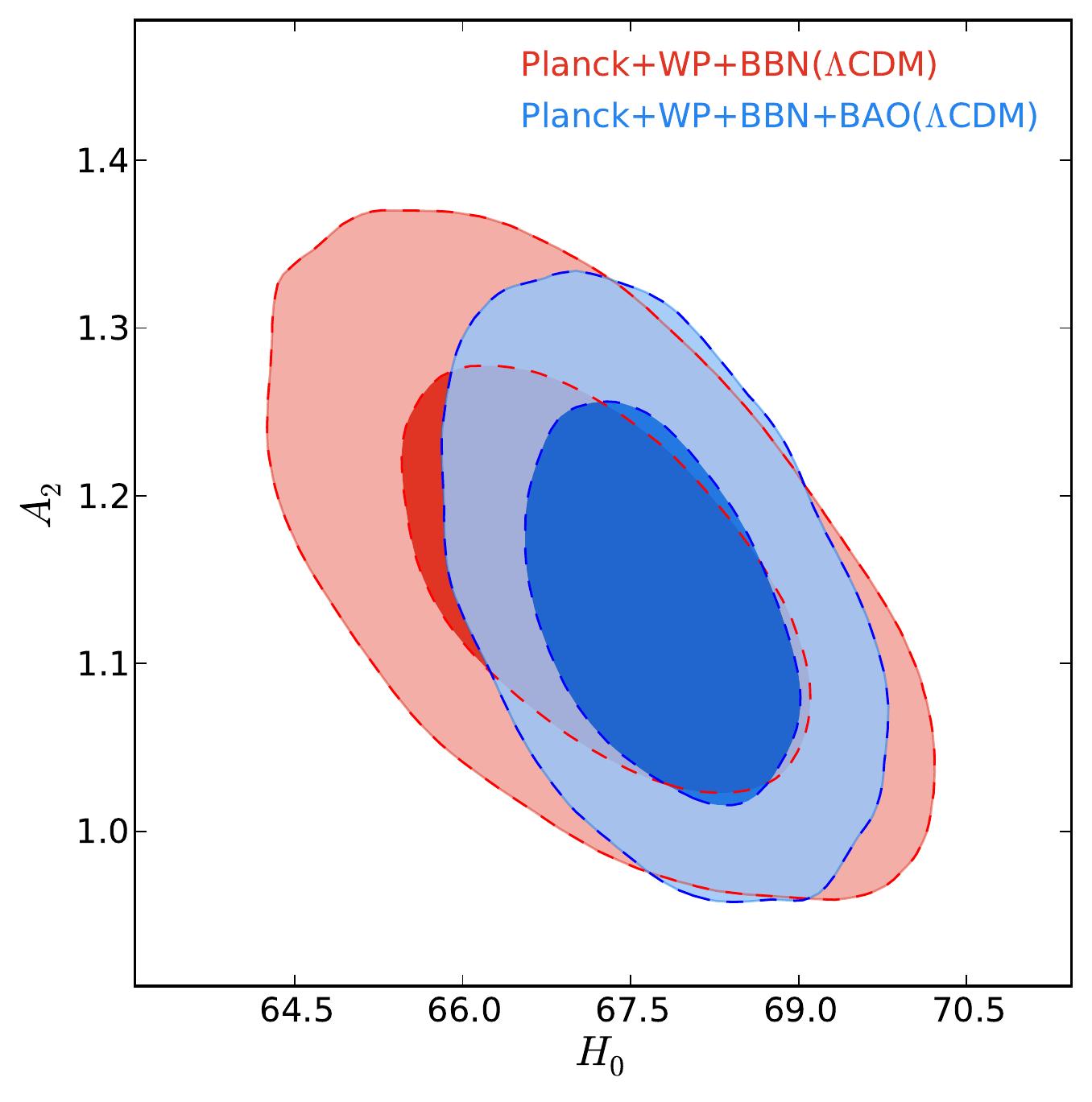}
\caption{\label{fig2} 2-D contour plots in the $\Omega_bh^2$ vs. $A_2$ (top panel) and 
$H_0$ vs. $A_2$ (bottom panel) planes, showing preferred parameter regions at the $68 \%$ and $95 \%$ confidence levels
in the case of the minimal $\Lambda$CDM model.}
\end{figure}

As expected from the discussion of sections \ref{sec:intro} and \ref{sec:nuc}, we find that the data provides an indication for $A_2$ being greater than one, roughly at the level of two standard deviations, even when adding the BAO dataset. We can also check explicitly in Figure \ref{fig2} (top panel) that there is a clear anti-correlation between $A_2$ and $\Omega_b h^2$: in order to improve the agreement between Planck data and deuterium abundance measurements, one needs either a value of the nuclear rate rescaling factor $A_2$ higher than one, or a value of the baryon density larger than the Planck mean value. This is could be expected, since deuterium is a decreasing function of both the $R_2$ rate and the baryon density $\Omega_b$. The lower panel of Figure \ref{fig2} also shows an interesting correlation between $A_2$ and the Hubble constant $H_0$. Letting $A_2$ vary yields a lower value for the Hubble constant in a combined Planck+WP+BBN analysis.

 \begin{table*}
\begin{center}
\begin{tabular}{|c|c|c|c|}
\hline\hline
Parameter & Planck+WP & Planck+WP & Planck+WP\\
 & +BBN & +BBN+HST & +BBN+BAO\\
\hline
$\Omega_bh^2$ &$0.02241\pm0.00042$ &$0.02261\pm0.00031$ &$0.02233\pm0.00029$\\
$\Omega_{\rm c}h^2$ &$0.1263\pm0.0055$ &$0.1281\pm0.0049$ &$0.1251\pm0.0051$\\
$\tau$ &$0.096\pm0.015$ &$0.099\pm0.014$ &$0.094\pm0.013$\\
$n_s$ &$0.979\pm0.017$ &$0.988\pm0.011$ &$0.974\pm0.010$ \\
$\log[10^{10} A_s]$ &$3.117\pm0.034$ &$3.128\pm0.030$ &$3.109\pm0.029$\\
$H_0 [\mathrm{km}/\mathrm{s}/\mathrm{Mpc}]$& $71.0\pm3.2$ &$72.8\pm2.0$ &$70.1\pm1.9$\\
$N_{\rm eff}$ &$3.56\pm0.40$ &$3.76\pm0.27$ &$3.43\pm0.30$\\
$A_2$ &$1.29\pm0.15$ &$1.33\pm0.14$ &$1.26\pm0.14$\\

\hline\hline
 \end{tabular}
 \caption{\label{tab2}Constraints on cosmological parameters (at the $68 \%$ confidence level) in the case of the extended $\Lambda$CDM model with extra relativistic degrees of freedom.}
 \end{center}
 \end{table*}

\begin{figure}
\includegraphics[width=.35\textwidth]{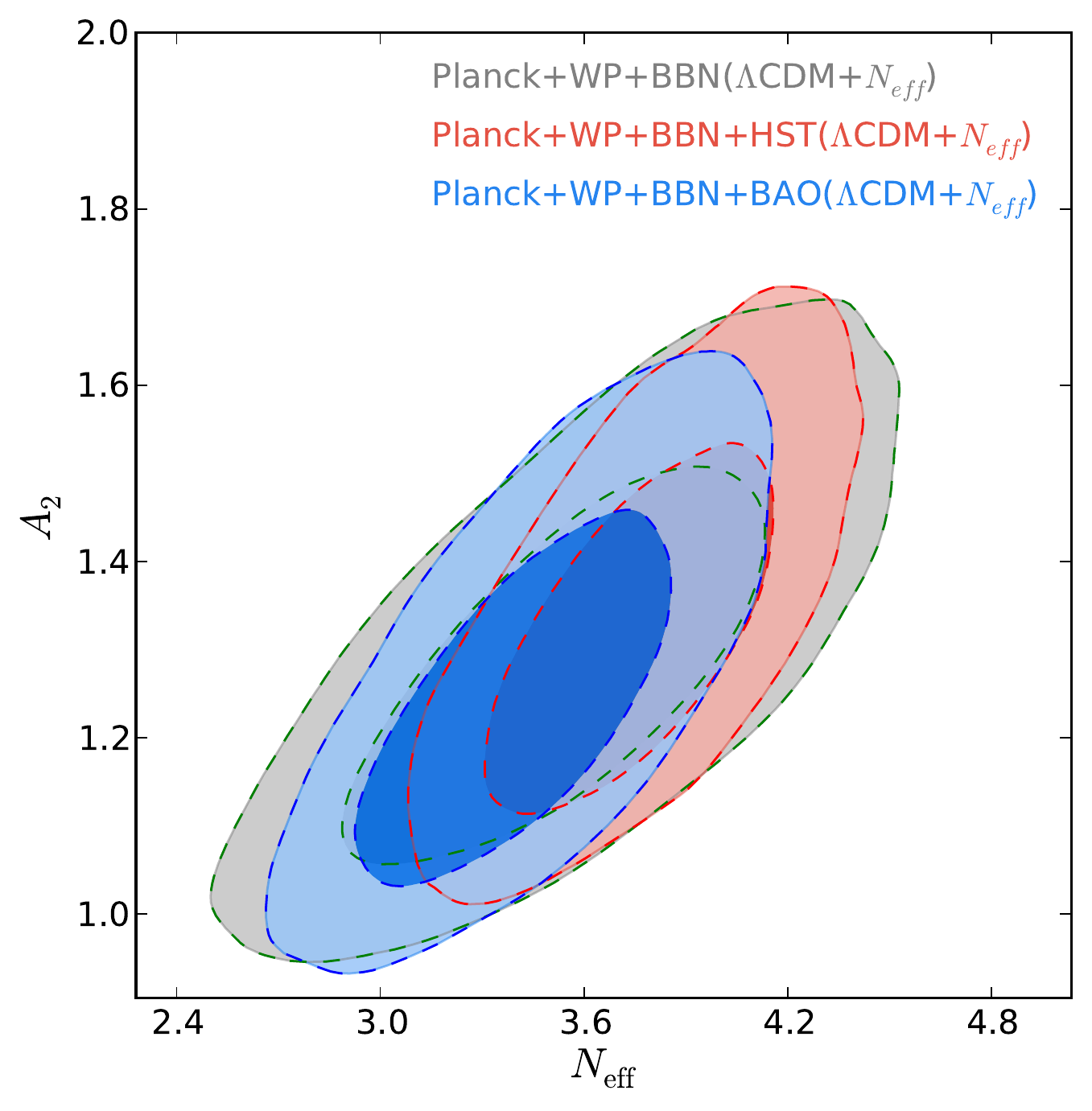}
\caption{\label{fig3} 2-D contour plots in the $N_{\rm eff}$ vs $A_2$ 
plane, showing preferred parameter regions at the $68 \%$ and $95 \%$ confidence levels
in the case of the extended $\Lambda$CDM model with extra relativistic degrees of freedom.}
\end{figure}

Given the fact that our results depend on the underlying cosmological model, it is interesting to investigate whether extensions of the standard $\Lambda$CDM model could bring the value of $A_2$ back in better agreement with the current experimental determination of $R_2(T)$ (corresponding by definition to $A_2=1$).

In Table~\ref{tab2}, we report the constraints when a variation in the neutrino effective number $N_{\rm eff}$ is allowed (to account, e.g., for extra relativistic degrees of freedom, or for non-standard physics in the neutrino sector). Even in that case, we can see that the combined Planck+WP+BBN and Planck+WP+BBN+BAO analyses show a preference for $A_2 >1$ at roughly the 2$\sigma$ level, even if the central value and error bar for $A_2$ are almost doubled. When the direct measurement of the Hubble parameter is included (case Planck+WP+BBN+HST), the indication for $A_2 >1$ is even stronger, at the 2.5$\sigma$ level. We can conclude that the preference for a large $d(p,\gamma)^3$He reaction rate is robust against the extension of the minimal cosmological model to a free $N_\mathrm{eff}$.

It is interesting to note that in Table~\ref{tab2}, the preferred value for the neutrino effective number $N_\mathrm{eff}$ is always
larger than the standard value $3.046$. As reported in section 6.4.4. of Ref.~\cite{PlanckXVI}, the ``standard'' Planck+WP+BBN analysis (assuming $A_2=1$) gives $N_\mathrm{eff}=3.02\pm0.27$~(68\% C.L.), while the CMB only result is $N_\mathrm{eff}=3.36\pm0.34$ (to be precise, in these results, the CMB dataset includes high-$\ell$ data from ACT and SPT, but the same trend is observed with only Planck+WP). With the present analysis, it becomes clear that this shift of $N_\mathrm{eff}$ towards its standard value is mostly  driven by the low experimental value of $R_2$. When $A_2$ is let free, the preference for $N_\mathrm{eff}>3.046$ persists even when deuterium measurements are  included. This can also be checked in Fig. \ref{fig3}, where we report the two dimensional likelihood contours in the $N_{eff}$ vs. $A_2$ plane for the three different datasets: Planck+WP+BBN, Planck+WP+BBN+HST, and Planck+WP+BBN+BAO. A correlation between $A_2$ and $N_\mathrm{eff}$ is clearly present: large values of $A_2$ remain compatible with Planck+WP+BBN data, provided that at the same time $N_\mathrm{eff}$ is larger than three. Such considerations reinforce the motivations for future experimental campaign to collect further data on the $d(p,\gamma)^3$He cross section in the few hundred keV range. Notice that for $A_2=1.055$, corresponding to the theoretical result of 
~\cite{Marcucci:2005zc} a standard value of $N_\mathrm{eff}$ is allowed at 68\% C.L.. If experiments would confirm the theoretical result $R_2^{th}(T)$ in the BBN energy range, the overall agreement of CMB and BBN data for a standard number of relativistic degrees of freedom would improve with respect to the $A_2=1$ case. This does not hold if the HST measurement of $H_0$ is included in the analysis.

\begin{table*}
\begin{center}
\begin{tabular}{|c|c|c|c|c|}
\hline\hline
Parameter & Planck+WP+BBN & Planck+WP+BBN & Planck+WP+BBN & Planck+WP+BBN \\
\hline
$\Omega_bh^2$ &$0.02242\pm0.00035$ &$0.02301\pm0.00051$ &$0.02227\pm0.00032$ &$0.02261\pm0.00042$ \\
$\Omega_{\rm c}h^2$ &$0.1169\pm0.0030$ &$0.1245\pm0.0055$ &$0.1185\pm0.0027$ &$0.1241\pm0.0053$\\
$\theta$ &$1.04179\pm0.00067$ &$1.04112\pm0.00078$ &$1.04153\pm0.00065$ &$1.04104\pm0.00079$ \\
$\tau$ &$0.087\pm0.013$ &$0.094\pm0.015$ &$0.087\pm0.013$ &$0.092\pm0.015$\\
$n_s$ &$0.9687\pm0.0085$ &$0.996\pm0.018$ &$0.9640\pm0.0075$ &$0.981\pm0.015$\\
$\log[10^{10} A_s]$ &$3.078\pm0.025$ &$3.111\pm0.034$ &$3.081\pm0.025$ &$3.105\pm0.033$\\
$H_0 [\mathrm{km}/\mathrm{s}/\mathrm{Mpc}]$& $68.8\pm1.4$ &$74.3\pm3.6$ & $56.7\pm5.4$ &$5905\pm6.4$ \\
$N_{\rm eff}$ &$[3.046]$ &$3.73\pm0.40$ &$[3.046]$ &$3.50\pm0.36$\\
$A_{\rm L}$ &$1.21\pm0.12$ &$1.25\pm0.13$ &$[1]$ &$[1]$\\
$\Omega_k$ &$[0]$ &$[0]$ &$-0.035\pm0.023$ &$-0.035\pm0.023$\\
$A_2$ &$1.067\pm0.086$ &$1.21\pm0.14$ &$1.100\pm0.084$ & $1.21\pm0.14$\\
\hline\hline
 \end{tabular}
 \caption{\label{tab3} Constraints on cosmological parameters (at the $68 \%$ confidence level) for several extensions of the $\Lambda$CDM model, with free parameters ($N_{\rm eff}$, $A_{\rm L}$, $\Omega_k$). We vary at most two of these extra parameters at the same time, and fix the other ones to their standard model value, indicated above between squared brackets.}
 \end{center}
 \end{table*}

\begin{figure}
\includegraphics[width=.35\textwidth]{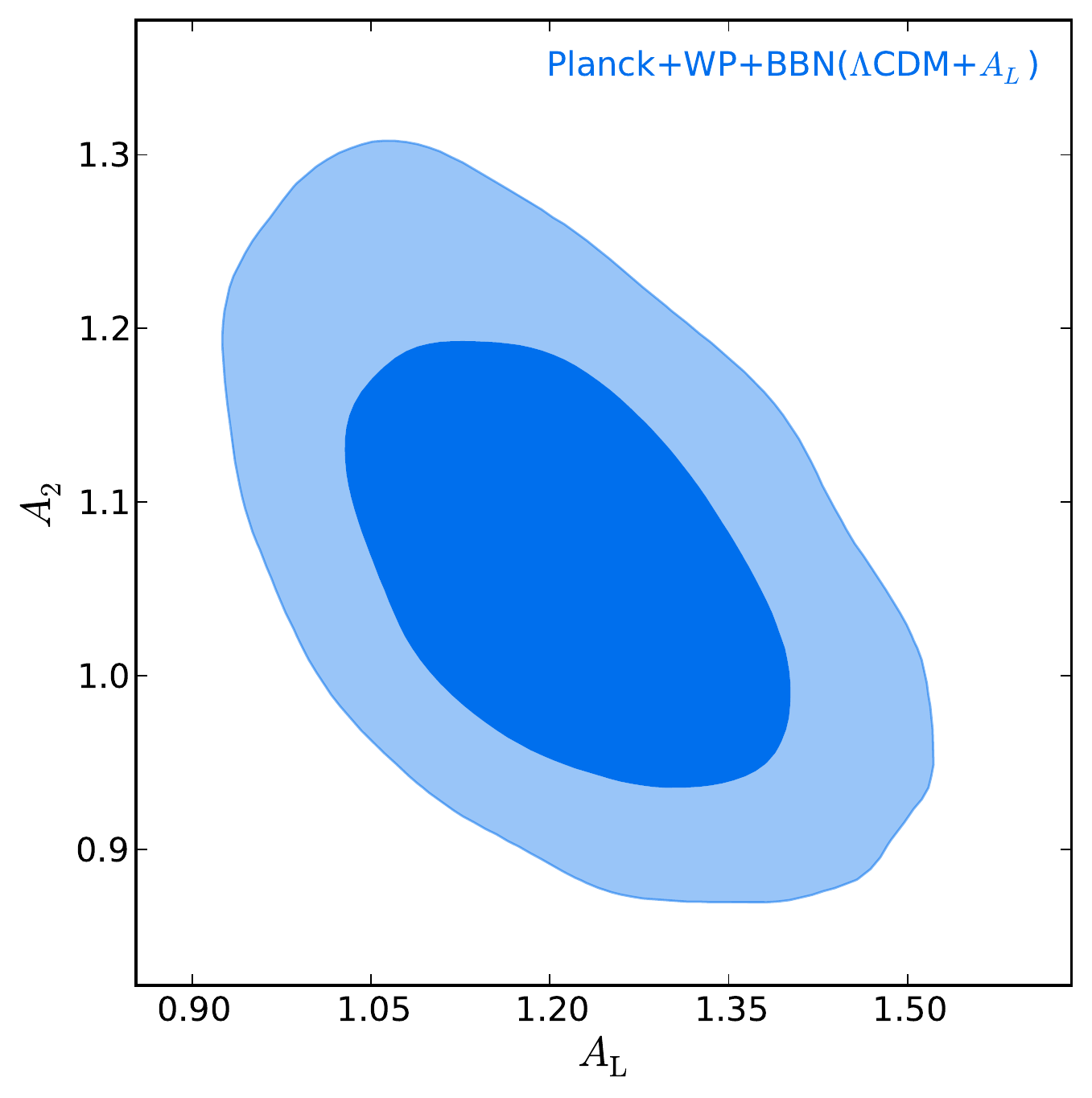}
\includegraphics[width=.35\textwidth]{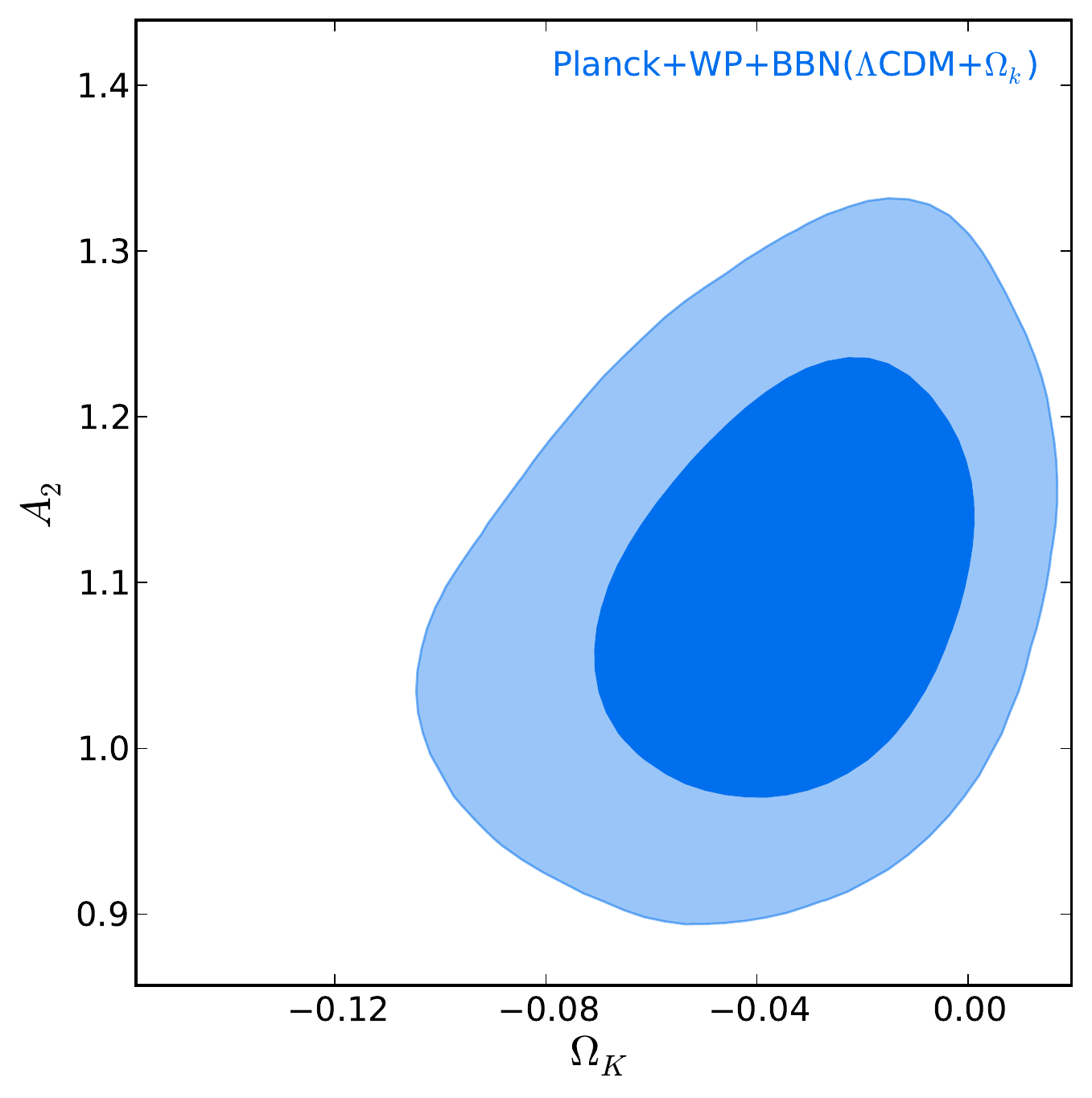}
\caption{\label{fig4} 2-D contour plots in the $A_{\rm L}$ vs $A_2$ (top panel) and 
$\Omega_k$ vs $A_2$ (bottom panel) planes showing probabilities at $68 \%$ and $95 \%$.}
\end{figure}

In Table \ref{tab3} we report the constraints on $A_2$ for further extensions of the minimal $\Lambda$CDM model, using the Planck+WP+BBN. We tried to vary the curvature parameter $\Omega_k$, despite the fact that $\Omega_k \neq 0$ is difficult to explain from a theoretical point of view, and almost excluded when BAO data is also included. With free spatial curvature and without BAO data, the evidence for $A_2>1$ is slightly weaker. Finally, we considered the case of a free CMB lensing amplitude parameter $A_{\rm L}$. Strictly speaking, this is not a physical extension of the $\Lambda$CDM model. The Planck data prefers $A_{\rm L}>1$, but as such, this result has no physical interpretation. It could be caused by a small and not yet identified systematic error affecting the Planck data (see the discussion in \cite{PlanckXVI}), or alternatively, it may account in some approximate way for a non-standard growth rate of large scale structures after recombination. We can see in Table \ref{tab3} that when $A_{\rm L}$ is left free, the $A_2$ parameter is well compatible with one. Our results for the joint confidence limits on $A_2$ vs. $\Omega_k$ and $A_2$ vs. $A_{\rm L}$ are shown in Fig.\ref{fig4}.

In summary, Planck+WP+BBN data consistently indicate that $A_2>1$ (suggesting a $d(p,\gamma)^3$He reaction rate closer to theoretical predictions than to experimental results) in the minimal $\Lambda$CDM model, as well as in a model with free $N_\mathrm{eff}$. The evidence for $A_2>1$ goes away when either $\Omega_k$ or $A_{\rm L}$ are promoted as free parameters (with $N_\mathrm{eff}=3.046$), but these scenarios are less theoretically motivated. Incidentally, Table \ref{tab3} also shows that with a free $\Omega_k$ or $A_{\rm L}$, and at the same time a free $N_\mathrm{eff}$, the evidence for $A_2>1$ persists.

\begin{table*}
\begin{center}
\begin{tabular}{|c|c|c|c|}
\hline\hline
Parameter & Planck+WP+ & Planck+WP+ & Planck+WP+ \\
&BICEP2+BBN&BICEP2+BBN&BICEP2+BBN \\
\hline
$\Omega_bh^2$ &$0.02209\pm0.00028$ &$0.02286\pm0.00044$ &$0.02236\pm0.00031$  \\
$\Omega_{\rm c}h^2$ &$0.1184\pm0.0027$ &$0.1300\pm0.0058$ &$0.1195\pm0.0027$ \\
$\theta$ &$1.04146\pm0.00063$ &$1.04050\pm0.00073$ &$1.04144\pm0.00063$  \\
$\tau$ &$0.088\pm0.012$ &$0.100\pm0.015$ &$0.101\pm0.015$ \\
$n_s$ &$0.9663\pm0.0072$ &$1.004\pm0.018$ &$0.9593\pm0.0080$ \\
$log[10^{10} A_s]$ &$3.082\pm0.024$ &$3.131\pm0.034$ &$3.115\pm0.031$ \\
$H_0 [\mathrm{km}/\mathrm{s}/\mathrm{Mpc}]$& $67.9\pm1.2$ &$75.5\pm3.7$ & $67.7\pm1.2$  \\
$r_{0.05}$ &$0.134\pm0.045 $ &$0.153\pm0.040$ &$0.163\pm0.040$ \\
$N_{\rm eff}$ &$[3.046]$ &$4.04\pm0.44$ &$[3.046]$ \\
$dn_s/dlnk$ &$[0]$ &$[0]$ &$-0.0256\pm0.0097$ \\
$A_2$ &$1.145\pm0.081$ &$1.40\pm0.17$ &$1.080\pm0.079$ \\
\hline\hline
 \end{tabular}
 \caption{Constraints on cosmological parameters (at the $68 \%$ confidence level) for the
 Planck+WP+BICEP2 dataset, with free parameters ($r_{0.05}$,$N_{\rm eff}$, $dn_s/dlnk$). We vary at most two of these extra parameters at the same time, and fix the other ones to their standard model value, indicated above between squared brackets.}
 \label{table4}
 \end{center}
 \end{table*}

\begin{figure}
\includegraphics[width=.35\textwidth]{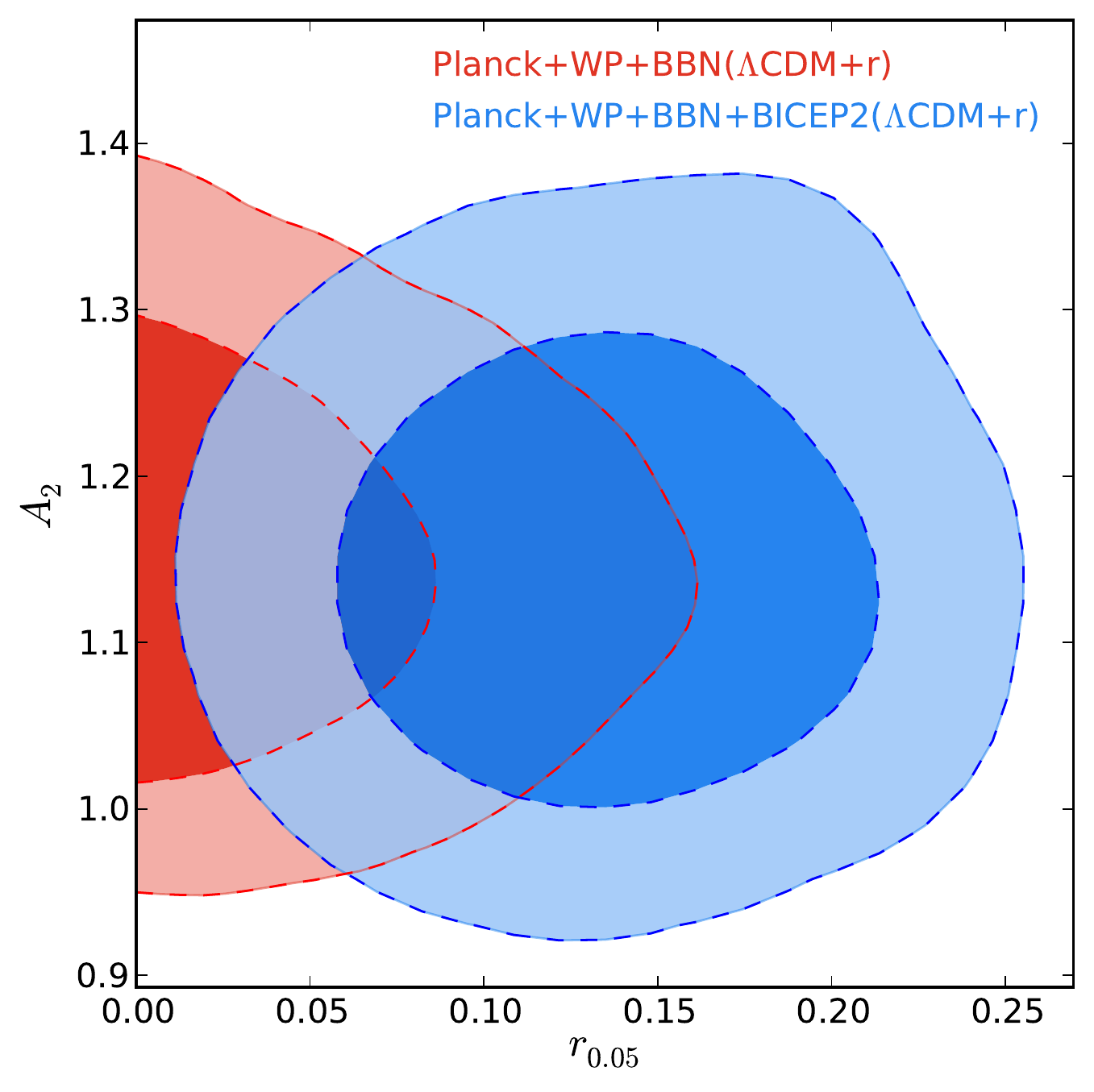}
\includegraphics[width=.35\textwidth]{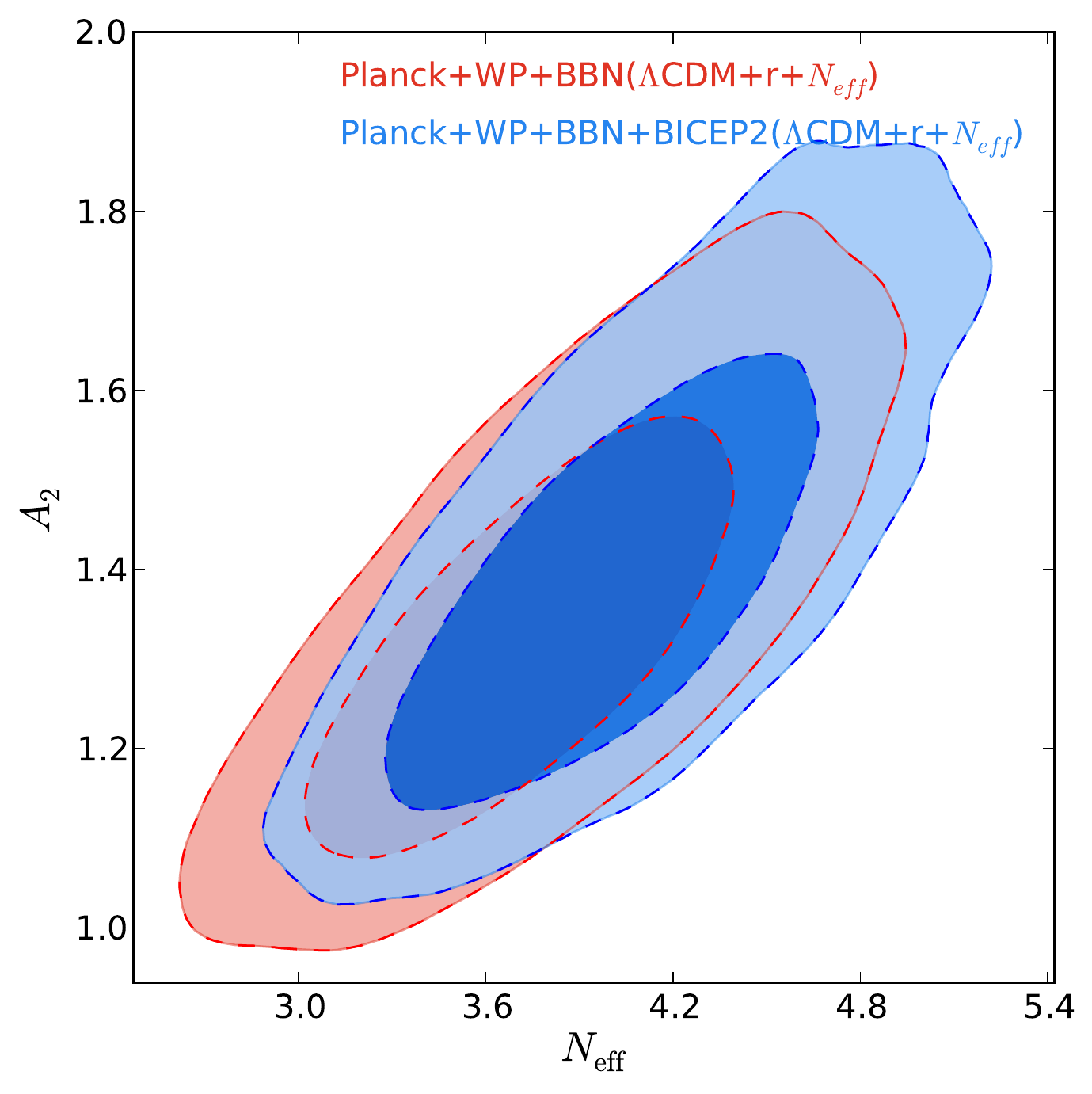}
\includegraphics[width=.35\textwidth]{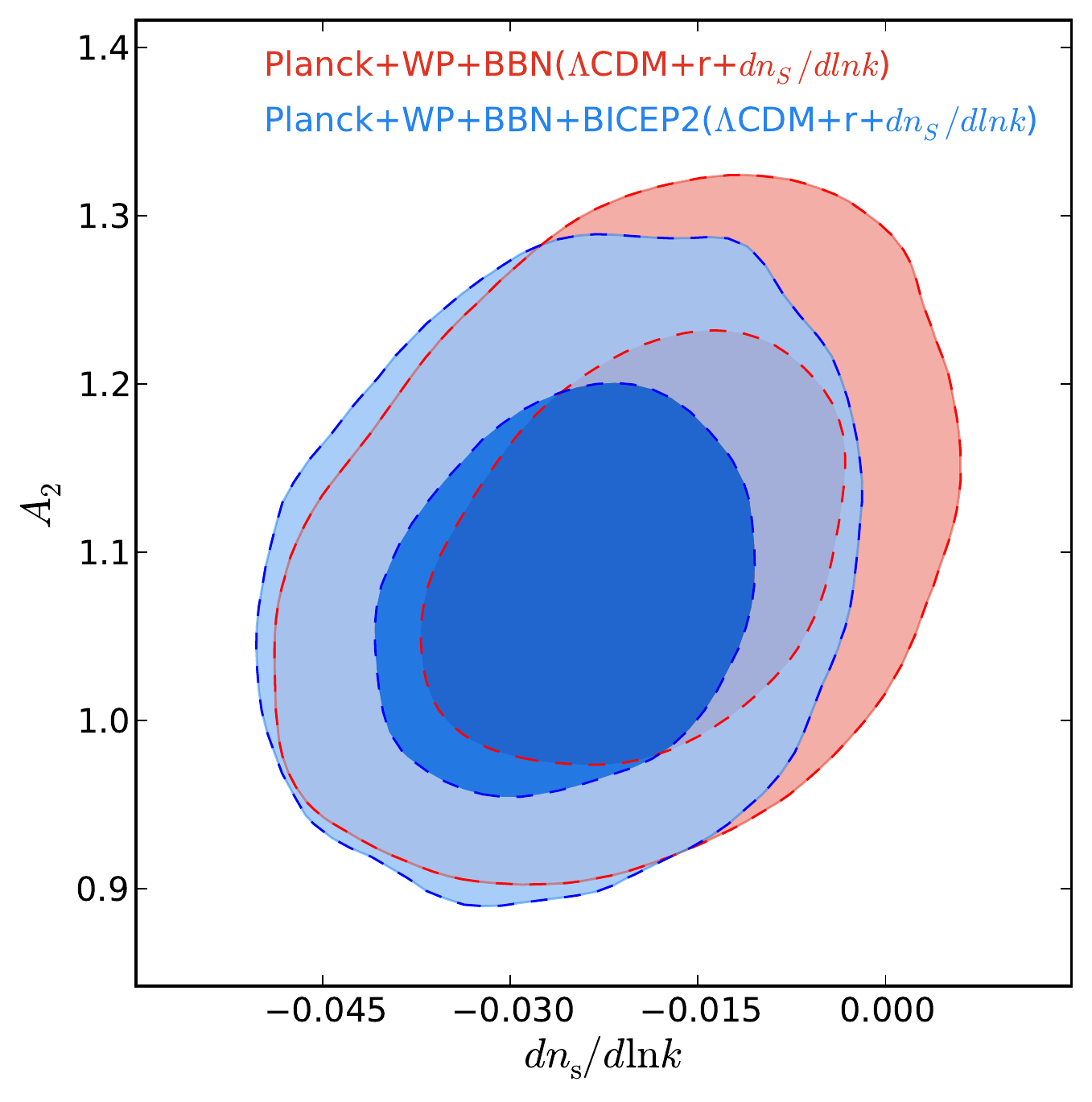}
\caption{\label{fig5} 2-D contour plots from the Planck+WP+BICEP2+BBN 
dataset in the $r$ vs $A_2$ (top panel), 
$N_{\rm eff}$ vs $A_2$ (center panel) and $dn_s/dlnk$ vs $A_2$ (bottom panel) 
planes showing probabilities at $68 \%$ and $95 \%$.}
\end{figure}

Finally, we have considered the Planck+WP+BICEP2+BBN dataset as stated in the previous
section. In Table \ref{table4} we report the constraints using this dataset, allowing for 
a gravitational wave background with tensor to scalar ratio $r_{0.05}$ at 
scales of $k=0.05 \,\rm{Mpc}^{-1}$. As we can see the indication for $A_2>1$ is still present in
this case. Allowing for a variation in $N_{\rm eff}$ provides even further evidence for
$A_2>1$ at more than two standard deviations. It is however interesting that when a running of
the primordial spectral index is considered, $A_2$ is now compatible with one in between one
standard devation. 
In \ref{fig5} we show the 2-D contour plots from the Planck+WP+BICEP2+BBN 
dataset in the $r_{0.05}$ vs $A_2$ (top panel), 
$N_{\rm eff}$ vs $A_2$ (center panel) and $dn_s/dlnk$ vs $A_2$ (bottom panel) 
planes showing probabilities at $68 \%$ and $95 \%$.
As we can see, while there is essentially no degeneracy between $A_2$ and $r_{0.05}$,
a degeneracy is clearly present between $A_2$ and $N_{\rm eff}$ and $dn_s/dlnk$.

In summary, the BICEP2 dataset, when combined with the
Planck data, provides an evidence either for a larger $N_{\rm eff}$,
either for a negative running of the spectral index $dn_s/dlnk$. 
In the first case a value of $A_2$ strictly larger than one is needed in 
order to be in agreement with BBN. In the second case, when
running is considered, $A_2$ is well compatible with one.
A precise measurement of $A_2$ from laboratory experiments could in principle help
in a significative way in discriminating between these two scenarios.

\section{Conclusions}\label{sec:conclusions}

In this work, we have shown that a combined analysis of Planck CMB data and of recent deuterium abundance measurements in metal-poor damped Lyman-alpha systems provides some piece of information on the radiative capture reaction $d(p,\gamma)^3$He, converting deuterium into helium. The value of the rate for this process represents the main source of uncertainty to date  in the BBN computation of the primordial deuterium abundance within a given cosmological scenario, parameterized by the baryon density $\Omega_b h^2$ and effective neutrino number $N_{\rm eff}$. The corresponding cross section has not been measured yet with a sufficiently low uncertainty and normalization errors in the BBN center of mass energy range, 30 - 300 keV. In addition to that, the best fit of available data appears to be systematically lower than the detailed theoretical calculation presented in~\cite{Marcucci:2005zc}. Both these issues should be addressed by performing new dedicated experimental campaigns. We think that an experiment such as LUNA at the underground Gran Sasso Laboratories may give an answer to this problem in a reasonably short time.

In fact, with the present underground $400~kV$ LUNA accelerator \cite{for} is possible to measure the $^2H(p,\gamma)^3He$ cross section in the $20<E_{cm}(keV)<260$ energy range with an accuracy  better than $3\%$, i.e. considerably better than the $9\%$ systematic uncertainty estimated in \cite{ma}. This goal can be achieved by using the large BGO detector already used in \cite{cas}.  This detector ensures a detection efficiency of about $70\%$ and a large angular coverage for the photons emitted by the $^2H(p,\gamma)^3He$ reaction. The accurate measurement of the $^2H(p,\gamma)^3He$ absolute cross section may be accomplished with the study of the angular distribution of emitted $\gamma$-rays by means of a large Ge(Li) detector \cite{and, an1}, in order to compare the data with "ab initio" modeling.

Our study shows that, interestingly, the combined analysis of Planck and deuterium abundance data returns a larger rate $A_2$ for this reaction than the best fit computed in \cite{Adelberger:2010qa}, where the authors exploit the available experimental information on $d(p,\gamma)^3$He cross section. On the other hand Planck is in better agreement with {\it ab initio} theoretical calculations. More precisely, when the reaction rate $A_2$ is chosen to match its present determination, Planck predicts a value of the primordial deuterium abundance in 2$\sigma$ tension with its direct astrophysical determination. When the same reaction rate $A_2$ is assumed instead to match theoretical calculations, the two values of the primordial deuterium abundance agree at the 1$\sigma$ level. We have shown that this conclusion holds  in the minimal $\Lambda$CDM cosmological model, as well as when allowing for a free effective neutrino number. In the latter case, the global likelihood analysis of astrophysical and cosmological data shows a direct correlation between $A_2$ and $N_{\rm eff}$, so that higher values for $A_2$ are in better agreement with non standard scenarios with extra relativistic degrees of freedom. 

Finally, we have shown that the inclusion of the new BICEP2 dataset also points towards a 
larger value for $A_2$, especially when $N_{\rm eff}$ is left free to vary. However, a running of the spectral index
could bring the value of $A_2$ back in agreement with one even when the BICEP2 dataset
is considered.

New experimental data on the $d(p,\gamma)^3$He reaction rate will therefore have a significant impact on the knowledge of  $N_{\rm eff}$ and of $dn_s/dlnk$ as well.

\subsection*{Acknowledgements}
We are pleased to thank  L.~E.~Marcucci, who kindly provided the results of the {\it ab initio} calculations of the $d(p,\gamma)^3 \mbox{He}$ astrophysical factor described in \cite{Marcucci:2005zc}
for discussions and help.

We thank the Planck Editorial Board for the internal refereeing of this work.


\begin{thebibliography}{99}


\bibitem{Iocco1} 
  F.~Iocco, G.~Mangano, G.~Miele, O.~Pisanti and P.~D.~Serpico,
  Phys.\ Rept.\  {\bf 472}, 1 (2009)
  [arXiv:0809.0631 [astro-ph]].

\bibitem{PlanckXVI} 
  P.~A.~R.~Ade {\it et al.}  [Planck Collaboration],
  arXiv:1303.5076 [astro-ph.CO].

  
\bibitem{parthenope} 
  O.~Pisanti, A.~Cirillo, S.~Esposito, F.~Iocco, G.~Mangano, G.~Miele and P.~D.~Serpico,
  Comput.\ Phys.\ Commun.\  {\bf 178}, 956 (2008)
  [arXiv:0705.0290 [astro-ph]].

  
\bibitem{Cooke:2013cba} 
  R.~Cooke, M.~Pettini, R.~A.~Jorgenson, M.~T.~Murphy and C.~C.~Steidel,
  arXiv:1308.3240 [astro-ph.CO].
  
\bibitem{Pettini-Cooke} 
  M.~Pettini and R.~Cooke,
  Mon.\ Not.\ Roy.\ Astron.\ Soc.\  {\bf 425}, 2477 (2012)
  [arXiv:1205.3785 [astro-ph.CO]].

 \bibitem{Adelberger:2010qa} 
  E.~G.~Adelberger, A.~B.~Balantekin, D.~Bemmerer, C.~A.~Bertulani, J.~-W.~Chen, H.~Costantini, M.~Couder and R.~Cyburt {\it et al.},
  Rev.\ Mod.\ Phys.\  {\bf 83}, 195 (2011)
  [arXiv:1004.2318 [nucl-ex]].
  
 \bibitem{casella}
 C. Casella {\it et al.} [LUNA Collaboration]
 Nucl. Phys. A {\bf 706}, 203 )2002)

\bibitem{Viviani:1999us} 
  M.~Viviani, A.~Kievsky, L.~E.~Marcucci, S.~Rosati and R.~Schiavilla,
  Phys.\ Rev.\ C {\bf 61}, 064001 (2000)
  [nucl-th/9911051].

\bibitem{Marcucci:2004sq} 
  L.~E.~Marcucci, K.~M.~Nollett, R.~Schiavilla and R.~B.~Wiringa,
  Nucl.\ Phys.\ A {\bf 777}, 111 (2006)
  [nucl-th/0402078].
  
\bibitem{Marcucci:2005zc} 
  L.~E.~Marcucci, M.~Viviani, R.~Schiavilla, A.~Kievsky and S.~Rosati,
  Phys.\ Rev.\ C {\bf 72}, 014001 (2005)
  [nucl-th/0502048].
  
  
 \bibitem{Nollett:2011aa} 
  K.~M.~Nollett and G.~P.~Holder,
  arXiv:1112.2683 [astro-ph.CO].
 
  \bibitem{Serpico:2004gx} 
  P.~D.~Serpico, S.~Esposito, F.~Iocco, G.~Mangano, G.~Miele and O.~Pisanti,
  JCAP {\bf 0412}, 010 (2004)
  [astro-ph/0408076].
  
  \bibitem{Rupak} G. Rupak, Nucl. Phys. A {\bf  678}, 409 (2000)
  
  \bibitem{Chen} J.W. Chen and M.J. Savage, Phys. Rev. C {\bf 60}, 60205 (1999)
    
  \bibitem{Leonard} D.S. Leonard {\it et al.}, Phys. Rev. C {\bf 73}, 045801 (2006) [nucl-ex/0601035]
  
\bibitem{Mangano} 
  G.~Mangano, G.~Miele, S.~Pastor, T.~Pinto, O.~Pisanti and P.~D.~Serpico,
  Nucl.\ Phys.\ B {\bf 729}, 221 (2005)
  [hep-ph/0506164].
  

\bibitem{PlanckXV} 
  P.~A.~R.~Ade {\it et al.}  [Planck Collaboration],
  arXiv:1303.5075 [astro-ph.CO].

\bibitem{bicep2}
  P.~A.~R.~Ade {\it et al.}  [BICEP2 Collaboration],
  arXiv:1403.3985 [astro-ph.CO].

\bibitem{hst}
  A.~G.~Riess, L.~Macri, S.~Casertano, H.~Lampeitl, H.~C.~Ferguson, A.~V.~Filippenko, S.~W.~Jha and W.~Li {\it et al.},
  Astrophys.\ J.\  {\bf 730} (2011) 119
   [Erratum-ibid.\  {\bf 732} (2011) 129]
  [arXiv:1103.2976 [astro-ph.CO]].

\bibitem{sdss-dr7}
  N.~Padmanabhan, X.~Xu, D.~J.~Eisenstein, R.~Scalzo, A.~J.~Cuesta, K.~T.~Mehta and E.~Kazin,
  Mon.\ Not.\ Roy.\ Astron.\ Soc.\  {\bf 427} (2012) 3,  2132
  [arXiv:1202.0090 [astro-ph.CO]].

\bibitem{sdss-dr9}
  L.~Anderson, E.~Aubourg, S.~Bailey, D.~Bizyaev, M.~Blanton, A.~S.~Bolton, J.~Brinkmann and J.~R.~Brownstein {\it et al.},
  Mon.\ Not.\ Roy.\ Astron.\ Soc.\  {\bf 427} (2013) 4,  3435
  [arXiv:1203.6594 [astro-ph.CO]].

\bibitem{wiggle-z}
  C.~Blake, T.~Davis, G.~Poole, D.~Parkinson, S.~Brough, M.~Colless, C.~Contreras and W.~Couch {\it et al.},
  Mon.\ Not.\ Roy.\ Astron.\ Soc.\  {\bf 415} (2011) 2892
  [arXiv:1105.2862 [astro-ph.CO]].
  
\bibitem{Lewis:2002ah} 
  A.~Lewis and S.~Bridle,
  Phys.\ Rev.\ D {\bf 66}, 103511 (2002)
  [astro-ph/0205436].
  
\bibitem{Audren:2012wb} 
  B.~Audren, J.~Lesgourgues, K.~Benabed and S.~Prunet,
  JCAP {\bf 1302}, 001 (2013)
  [arXiv:1210.7183 [astro-ph.CO]].
  
\bibitem{Lewis:2013hha} 
  A.~Lewis,
  arXiv:1304.4473 [astro-ph.CO].


\bibitem{Calabrese} 
  E.~Calabrese, A.~Slosar, A.~Melchiorri, G.~F.~Smoot and O.~Zahn,
  Phys.\ Rev.\ D {\bf 77}, 123531 (2008)
  [arXiv:0803.2309 [astro-ph]].

\bibitem{giusarma14}
  E.~Giusarma, E.~Di Valentino, M.~Lattanzi, A.~Melchiorri and O.~Mena,
  arXiv:1403.4852 [astro-ph.CO];
J.~-F.~Zhang, Y.~-H.~Li and X.~Zhang,
  arXiv:1403.7028 [astro-ph.CO];
C.~Dvorkin, M.~Wyman, D.~H.~Rudd and W.~Hu,
  arXiv:1403.8049 [astro-ph.CO].

\bibitem{for} A. Formicola et al. (LUNA collaboration), Nucl. Instr. and Meth. A 507 (2003) 609.
\bibitem{ma} L. Ma et al., Phys. Rev. C 55, 588 (1997).
\bibitem{cas} C. Casella et al. (LUNA collaboration): Nuclear Physics A 706 (2002) 203�216.
\bibitem{and} M. Anders et al. (LUNA collaboration): Eur. Phys. J. A (2013) 49: 28.
\bibitem{an1} M. Anders et al. (LUNA collaboration): Phys. Rev. Lett., submitted.

\end{thebibliography}
\end{document}